\documentclass{PoS}

\PoS{PoS(LAT2005)337}

\usepackage[dvips]{graphicx}
\usepackage{amsmath}

\newcommand{\be}{\begin{equation}}
\newcommand{\ee}{\end{equation}}
\newcommand{\bea}{\begin{eqnarray}}
\newcommand{\eea}{\end{eqnarray}}

\newcommand{\bt}{\begin{tabbing}}
\newcommand{\et}{\end{tabbing}}
\newcommand{\bi}{\begin{itemize}}
\newcommand{\ei}{\end{itemize}}
\newcommand{\ben}{\begin{enumerate}}
\newcommand{\een}{\end{enumerate}}

\title{%
\begin{picture}(0,0)(0,0)%
\put(355,75){\makebox(0,0)[l]{\textnormal{\normalsize KEK-CP-174}}}%
\put(355,60){\makebox(0,0)[l]{\textnormal{\normalsize BNL-HET-05/23}}}%
\end{picture}%
$K_{l3}$ form factor with two-flavors of dynamical domain-wall quarks}

\ShortTitle{$K_{l3}$ form factor with two flavors of dynamical domain-wall quarks}

\author{RBC Collaboration:}

\author{C.~Dawson
        \\
        RIKEN-BNL Research Center, Brookhaven National Laboratory, 
        Upton, NY 11973, USA}

\author{T.~Izubuchi
        \\
        Institute of Theoretical Physics, Kanazawa University, 
        Ishikawa 920-1192, Japan
        \\
        RIKEN-BNL Research Center, Brookhaven National Laboratory, 
        Upton, NY 11973, USA}

\author{\speaker{T.~Kaneko}
        \\
        High Energy Accelerator Research Organization (KEK),
        Ibaraki 305-0801, Japan
        \\
        The Graduate University for Advanced Studies,
        Ibaraki 305-0801, Japan}

\author{S.~Sasaki
        \\
        Department of Physics, University of Tokyo, Tokyo 113-0033, Japan 
        \\
        RIKEN-BNL Research Center, Brookhaven National Laboratory, 
        Upton, NY 11973, USA}

\author{A.~Soni
        \\
        Physics Department, Brookhaven National Laboratory, 
        Upton, NY 11973, USA}

\abstract{
We report on our calculation of $K \to \pi$ vector form factor
by numerical simulations of two-flavor QCD 
on a $16^3 \times 32 \times 12$ lattice at $a \simeq 0.12$~fm
using domain-wall quarks and DBW2 glue.
Our preliminary result at a single sea quark mass 
correponding to $m_{\rm PS}/m_{\rm V} \simeq 0.53$ 
shows a good agreement with previous estimate in quenched QCD
and that from a phenomenological model.
}

\FullConference{XXIIIrd International Symposium on Lattice Field Theory\\
                 25-30 July 2005\\
                 Trinity College, Dublin, Ireland}

\begin{document}

\section{Introduction}

Currently, $K \to \pi l \nu_l$ ($K_{l3}$) decay provides 
the most precise determination
of the Cabibbo-Kobayashi-Maskawa (CKM) matrix element $|V_{us}|$ 
through $\Gamma \propto |V_{us}|^2 |f_+(0)|^2 $, where 
$\Gamma$ is the decay rate and $f_+(q^2)$ is the form factor 
defined from the $K \to \pi$ matrix element of the weak vector current
\bea
   \left\langle 
      \pi(p^{\prime}) \left| V_\mu \right| K(p)
   \right\rangle
   & = & 
      ( p_{\mu} + p^{\prime}_{\mu} )\, f_+(q^2) 
     +( p_{\mu} - p^{\prime}_{\mu} )\, f_-(q^2),
   \hspace{5mm}
   q^2 = (p-p^{\prime})^2.
   \label{eqn:f+-}
\eea
While the PDG values for the CKM matrix elements~\cite{PDG2004} 
show a $2\sigma$ deviation from the CKM unitarity
\bea
   |V_{ud}|^2 + |V_{us}|^2 + |V_{ub}|^2 = 1 - \delta,
   \hspace{5mm}
   \delta = 0.0033(15),
   \label{eqn:unitarity}
\eea
recent experiments for $\Gamma$~\cite{E865,KTeV,NA48,KLOE} prefer 
a slightly larger value of $|V_{us}|$,
which is consistent with the unitarity ($\delta\!=\!0$).
However, 
in order to make a definite conclusion on this issue,
$f_+(0)$ has to be determined theoretically with an accuracy of about 1\%.

In chiral perturbation theory (ChPT), $f_+(0)$ is expanded in powers of 
meson masses $M_K$, $M_\pi$, and $M_\eta$
\bea
   f_+(0) & = & 1 + f_2 + f_4 + \cdots,
   \hspace{5mm}
   (f_n = O(M_{K,\pi,\eta}^n)).
\eea
Thanks to the Ademollo-Gatto theorem~\cite{Ademollo-Gatto}, 
which states that $SU(3)$ breaking effects in $f_+(0)$ start 
at $O((m_s-m_{ud})^2)$,
poorly-known low-energy constants of the chiral Lagrangian 
do not enter the ChPT formula for $f_2$~\cite{f2},
and hence it is precisely determined as $-0.023$.
However, this is not the case for the higher order corrections
$f_n$ $(n \! \geq \!4$)~\cite{f4:ChPT}.
Therefore, a phenomenological estimate of $f_4$ 
based on the quark model~\cite{f4:LR} has been used 
in previous analyses of $|V_{us}|$.

In order to estimate $f_+(0)$
without relying on any phenomenological model,
lattice calculations have been carried out 
first in the quenched approximation~\cite{fn:Nf0:italy},
and later in unquenched QCD~\cite{fn:Nf2:JLQCD,fn:Nf3:FNAL}.
In these calculations, however,
conventional Wilson- or Kogut-Susskind-type quark actions are employed,
and hence chiral properties of $f_+(0)$ may be significantly 
affected by the explicit breaking of chiral or flavor symmetry.
Since the ChPT formula for $f_2$ plays a crucial role 
in the chiral extrapolation of lattice data,
it is advantageous to use a quark action,
which posses chiral symmetry even at finite lattice spacing.
In this work, we calculate $f_+(0)$ in two-flavor dynamical QCD
using the DBW2 gauge~\cite{DBW2} and the domain-wall quark
actions~\cite{DWF},
with which the hadron spectrum and the kaon $B$ parameter
show good chiral properties~\cite{Spectrum+BK:Nf2QCD:RBC}.

\section{Simulation method}


Our calculations are carried out on a $16^3 \times 32$ lattice 
with statistics of 4750 HMC trajectories.
While we simulate a single value for the lattice spacing 
$a^{-1}\!=\!$1.69(5) GeV,
we expect that the scaling violation in $f_+$ is not large,
since the employed lattice action is (automatically) $O(a)$-improved.
The size of the fifth dimension is 
set to $L_5\!=\!12$, 
which leads to the residual mass of a few MeV.
We refer to Ref.~\cite{Spectrum+BK:Nf2QCD:RBC} for details on 
the gauge ensembles used in this study.


At the moment, we complete our calculation at 
a single sea quark mass $m_{ud}\!=\!0.02$,
which is roughly half the physical strange quark mass.
%
%
Three heavier masses $m_{\rm s}\!=\!0.03$, 0.04 and 0.05 are employed 
for the valence strange quarks.

\section{Extraction of form factor}

\begin{figure}[b]
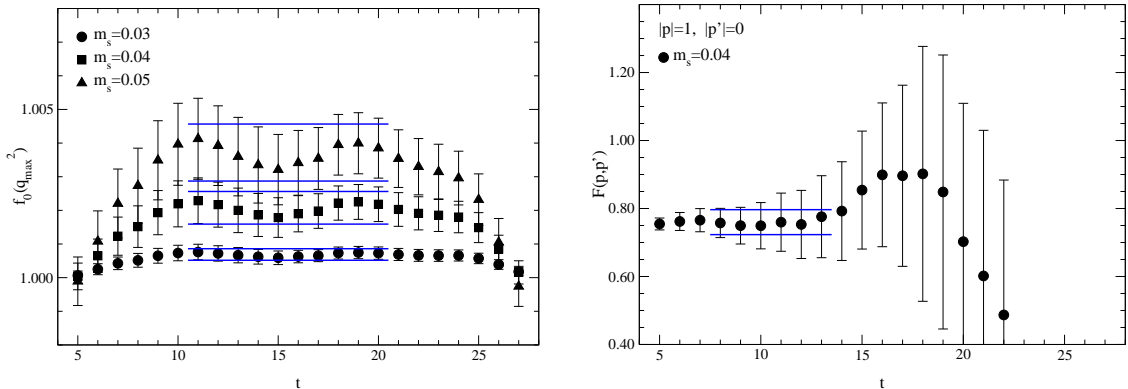

\vspace{2mm}
\begin{center}
\includegraphics[width=0.47\linewidth,clip]{kff_mud0_mom00.eps}
\hspace{5mm}
\includegraphics[width=0.47\linewidth,clip]{kff_mud0_mval02_mom10.eps}
\end{center}
\vspace{-7mm}
\caption{
Plots of $f_0(q_{\rm max}^2)$ (left figure) and 
$F(p,p^{\prime})$ with $|{\bf p}|\!=\!1$ and $|{\bf p}^{\prime}|\!=\!0$ (right figure)
as a function of $t$.
}
\label{fig:R_F}
\end{figure}

The so-called scalar form factor 
\bea 
   f_0(q^2)
   & = &
   f_+(q^2) + \frac{q^2}{M_K^2 - M_{\pi}^2} f_-(q^2)
   \label{eqn:f0}
\eea
at $q_{\rm max}^2\!=\!(M_K-M_\pi)^2$
can be extracted from the double ratio
proposed in Ref.~\cite{dble_rat},
\bea
   R(t,t^{\prime})
   & = &
   \frac{C_{4}^{K\pi}(t,t^{\prime};{\bf 0},{\bf 0}) \, 
         C_{4}^{\pi K}(t,t^{\prime};{\bf 0},{\bf 0})}
        {C_{4}^{KK}(t,t^{\prime};{\bf 0},{\bf 0}) \, 
         C_{4}^{\pi\pi}(t,t^{\prime};{\bf 0},{\bf 0})}
   \hspace{2mm} 
   \xrightarrow[t,(t^{\prime}-t) \to \infty]{}
   \hspace{2mm} 
   \frac{(M_K+M_\pi)^2}{4 M_K M_\pi} |f_0(q_{\rm max}^2)|^2,
   \\
   C_{\mu}^{PQ}(t,t^{\prime};{\bf p},{\bf p}^{\prime})
   & = &
   \sum_{{\bf x},\, {\bf x}^{\prime}}
   \left\langle 
      O_{Q}({\bf x}^{\prime},t^{\prime}) \,
      V_{\mu}({\bf x},t) \, 
      O_{P}^{\dagger}({\bf 0},0)
   \right\rangle
   \, 
   e^{-i{\bf p}^{\prime}({\bf x}^{\prime}-{\bf x})}
   e^{-i{\bf p}{\bf x}},
   \hspace{2mm}
   (P,Q = \pi \mbox{ or } K),
\eea
where $O_{\pi(K)}$ represents the interpolating operator for pion (kaon).
As shown in Fig.~\ref{fig:R_F},
$f_0(q_{\rm max}^2)$ is determined with an accuracy of $\lesssim$~0.1\,\%,
since various uncertainties of the three-point function, 
such as the statistical fluctuation,
are canceled at least partially in the ratio.

To study the $q^2$ dependence of the form factor,
we calculate 
\bea
   F(p,p^{\prime})
   & = &
   \frac{f_+(q^2)}{f_0(q_{\rm max}^2)}
   \left( 1 + \frac{E_K({\bf p})-E_\pi({\bf p}^{\prime})}
                   {E_K({\bf p})+E_\pi({\bf p}^{\prime})} \, \xi(q^2) \right),
   \hspace{5mm}
   \xi(q^2) = f_-(q^2)/f_+(q^2)
   \label{eq:F24}
\eea
from a ratio 
\bea
   \frac{C_{4}^{K\pi}(t,t^{\prime};{\bf p},{\bf p}^{\prime})\,
         C^{K}(t;{\bf 0})\,C^{\pi}(t^{\prime}-t;{\bf 0})}
        {C_{4}^{K\pi}(t,t^{\prime};{\bf 0},{\bf 0})\,
         C^{K}(t;{\bf p})\,C^{\pi}(t^{\prime}-t;{\bf p}^{\prime})}
   & \xrightarrow[t,(t^{\prime}-t) \to \infty]{} & 
   \frac{E_K({\bf p})+E_\pi({\bf p}^{\prime})}{M_K+M_\pi} \, 
   F(p,p^{\prime}),
\eea
where $C^{\pi(K)}(t;{\bf p})$ represents the pion (kaon) propagator
with the spatial momentum ${\bf p}$.
Figure~\ref{fig:R_F} shows a plot of $F(p,p^{\prime})$.
The accuracy of $F(p,p^{\prime})$ is typically 5\,--\,10\% 
for the spatial momenta $|{\bf p}|,|{\bf p}^{\prime}| \leq \sqrt{2}$.

\begin{figure}[b]
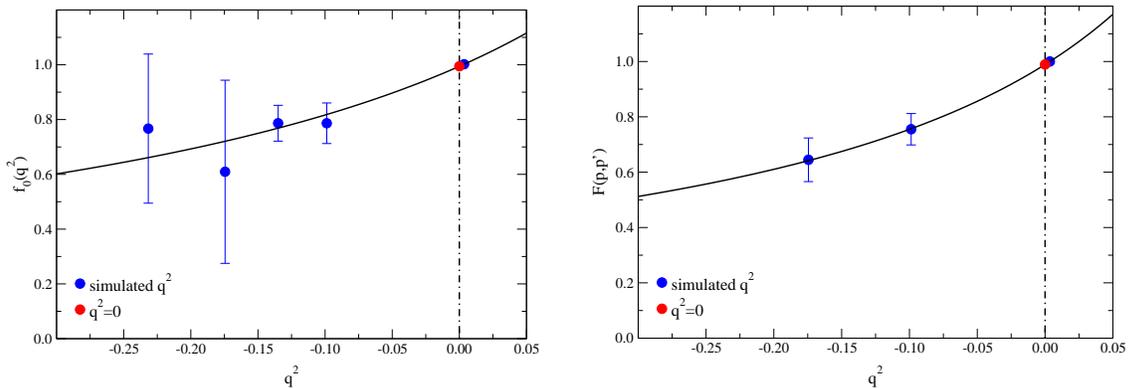

\vspace{4mm}
\begin{center}
\includegraphics[width=0.47\linewidth,clip]{f0_vs_q2_mud0_mval02.eps}
\hspace{5mm}
\includegraphics[width=0.47\linewidth,clip]{F24_vs_q2_mud0_mval02.eps}
\end{center}
\vspace{-7mm}
\caption{
Interpolation of $f_0(q^2)$ (left figure) and 
$F(p,p^{\prime})$ with ${\bf p}^\prime\!=\!{\bf 0}$ (right figure) 
to $q^2\!=\!0$.
Both figures show data at $m_s\!=\!0.04$.
}
\label{fig:q2interp}
\end{figure}

In order to convert $F(p,p^{\prime})$ to $f_0(q^2)$,
$\xi(q^2)$ is estimated from the double ratio
proposed in Ref.\cite{fn:Nf0:italy}
\bea
   &&
   R_k(t,t^{\prime};{\bf p},{\bf p}^{\prime})
   =
   \frac{C_k^{K\pi}(t,t^{\prime};{\bf p},{\bf p}^{\prime}) \,
         C_4^{KK}(t,t^{\prime};{\bf p},{\bf p}^{\prime})}
        {C_4^{K\pi}(t,t^{\prime};{\bf p},{\bf p}^{\prime}) \,
         C_k^{KK}(t,t^{\prime};{\bf p},{\bf p}^{\prime})}
   \hspace{5mm}
   (k=1,2,3),
   \\
   &&
   \xi(q^2)
   = 
   \frac{-(E_K({\bf p})\!+\!E_K({\bf p}^{\prime}))\,
          (p\!+\!p^{\prime})_k
         +(E_K({\bf p})\!+\!E_\pi({\bf p}^{\prime}))\,
          (p\!+\!p^{\prime})_k \, R_k}
        { (E_K({\bf p})\!+\!E_K({\bf p}^{\prime}))\,
          (p\!-\!p^{\prime})_k
         -(E_K({\bf p})\!-\!E_\pi({\bf p}^{\prime}))\,
          (p\!+\!p^{\prime})_k \, R_k}.
\eea
We observe that $\xi(q^2)$ has a mild dependence on the valence quark mass 
and its magnitude is typically $\simeq -0.01$ with 50\,--\,100\% error.

\section{Interpolation to $q^2\!=\!0$}

Here, we test two methods to determine $f_+(0)$ ($=\!f_0(0)$).
\bi
   \item method-1:
         As in Ref.\cite{fn:Nf0:italy}, we first calculate $f_0(q^2)$ 
         at each simulated $q^2$ from $f_0(q_{\rm max}^2)$,
         $F(p,p^{^\prime})$ and $\xi(q^2)$, 
         and then interpolate $f_0(q^2)$ to $q^2\!=\!0$.
   \item method-2:
         As in Ref.\cite{fn:Nf2:JLQCD}, $F(p,p^{\prime})$ 
         ($\xi(q^2)$) is interpolated (extrapolated) to $q^2\!=\!0$, 
         and then $f_0(0)$ is calculated at $q^2\!=\!0$.
         The interpolation of $F(p,p^{\prime})$ is carried out 
         using data with fixed $|{\bf p}|$ (or $|{\bf p}^{\prime}|$) so that
         we can unambiguously identify $|{\bf p}^{\prime}|$ $(|{\bf p}|)$ 
         corresponding to $q^2\!=\!0$, which is needed to convert 
         $F(p,p^{\prime})|_{q^2\!=\!0}$ to $f_0(0)$.
         We repeat this analysis for two data sets 
         with ${\bf p}\!=\!{\bf 0}$ and ${\bf p}^{\prime}\!=\!{\bf 0}$, 
         and take the average of results for $f_0(0)$.
\ei

We test quadratic and polar fits 
\bea
   {\mathcal O}(q^2) & = & {\mathcal O}(0) \cdot (1+c_1\, q^2 + c_2\,q^4),
   \\
   {\mathcal O}(q^2) & = & \frac{{\mathcal O}(0)}{1-c_1\, q^2}
\eea
for the interpolation of $f_0$ in method-1 and $F$ for method-2,
while only the quadratic fit is tested for $\xi$.
We observe that four possible ways (quadratic or polar fit for method-1 or 2) 
give mutually consistent results with an accuracy of $\lesssim$ 1\%.
This is because we have very accurate data of $f_0(q_{\rm max}^2)$ 
near $q^2\!=\!0$, and hence the uncertainty due to the choice of
the interpolation method is not large.

However, we observe that method-2 leads to a slightly smaller error
of $f_0(0)$ than method-1, since 
the kinematical factor $(E_K({\bf p})\!-\!E_\pi({\bf p}^{\prime}))/
(E_K({\bf p})\!+\!E_\pi({\bf p}^{\prime}))$ in Eq.~(\ref{eq:F24})
is not large at $q^2\!=\!0$,
and hence the uncertainty of $\xi(0)$ has small influence to $f_0(0)$
in method-2.
Therefore, we employ $f_+(0)$ obtained from method-2 with 
the quadratic fit in the following.

\section{Chiral extrapolation}

From the Ademollo-Gatto theorem, 
the higher order correction
\bea
   \Delta f = f_+(0) - (1+f_2) = \sum_{k=2}^{\infty} f_{2k}
\eea
is proportional to $(m_s\!-\!m_{ud})^2$.
Therefore, as in Ref.\cite{fn:Nf0:italy},
it is convenient to consider a ratio
\bea
   R_{\Delta f} = \frac{\Delta f}{(M_K^2-M_\pi^2)^2},
\eea
for the chiral extrapolation of $f_+(0)$.
In this analysis, $f_2$ at the simulated quark masses is calculated 
by using the ChPT formula in unquenched QCD~\cite{f2}\footnote{
Very recently, 
$f_2$ is calculated in partially quenched ChPT (PQChPT)~\cite{f2:pqQCD}.
We confirm that $f_+(0)$ obtained from the chiral extrapolation with 
the PQChPT formula for $f_2$ is consistent with the result presented 
in the text
within the statistical errors.
}.

\begin{figure}[b]
\vspace{2mm}
\begin{center}
\includegraphics[width=0.6\linewidth,clip]{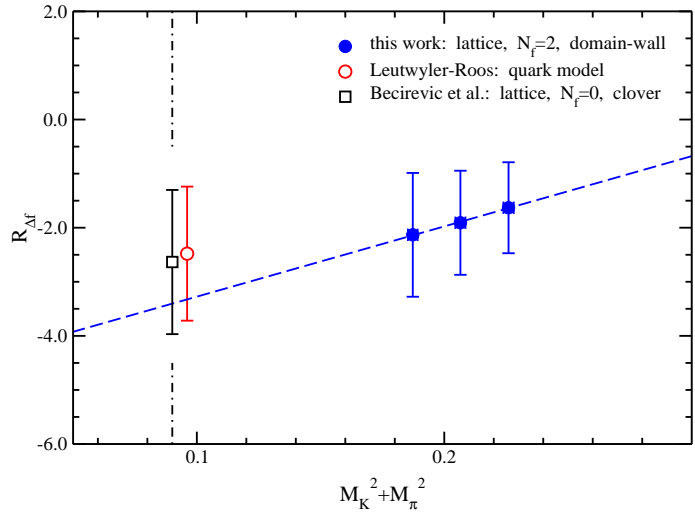}
\end{center}
\vspace{-7mm}
\caption{
Quark mass dependence of $R_{\Delta f}$. 
Open circle and square represent values corresponding to 
the phenomenological estimate~\cite{f4:LR}
and the quenched result~\cite{fn:Nf0:italy}.
}
\label{fig:df_vs_mq}
\end{figure}

Our results for $R_{\Delta f}$ are plotted in Fig.~\ref{fig:df_vs_mq}.
Since the single sea quark mass is simulated so far,
we extrapolate $R_{\Delta f}$ only in terms of the valence quark mass
by naively assuming that its sea quark mass dependence is not large
below the simulated sea quark mass.
From a simple linear fit shown in Fig.~\ref{fig:df_vs_mq},
we obtain $f_+(0)\!=\!0.955(12)$, 
which is consistent with both of the previous phenomenological 
estimate 0.961(8)~\cite{f4:LR} 
and the quenched result 0.960(9)~\cite{fn:Nf0:italy}.
We note that recent unquenched calculations 
in Refs.\cite{fn:Nf2:JLQCD,fn:Nf3:FNAL}
also obtained similar values for $f_+(0)$.

\section{Conclusions}
 
We have calculated $f_+(0)$ in two-flavor QCD using the domain-wall quarks.
Our preliminary result is consistent with the previous 
phenomenological estimate with an accuracy of 1\% level.
Our estimate combined with recent experimental results of $\Gamma$
leads to $|V_{us}|$ which is consistent with the CKM unitarity.
For instance, by using $\Gamma$ from the E865 Collaboration~\cite{E865},
we obtain $|V_{us}|\!=\!0.229(4)$ which leads to
\bea
   %
   |V_{ud}|^2 + |V_{us}|^2 + |V_{ub}|^2 = 1 - \delta,
   \hspace{5mm}
   \delta = -0.001(2).
   \label{eqn:unitarity:rst}
\eea
We note that, however, our result has an additional uncertainty 
arising from the fact that 
we have not taken the limit of the physical sea quark mass.
To remove this uncertainty,
our simulations at two different sea quark masses are in progress.

The nice consistency between our and previous quenched estimates of $f_+(0)$
may suggest that the systematic error due to the quenched
approximation for strange quarks is not large.
This point, however, has to be confirmed 
by extending our calculation to three-flavor QCD.

\hspace{5mm}

We thank RIKEN, BNL and the U.S. Department of Energy for providing 
the facilities essential for this work.
The work of TK is supported in part by the Grant-in-Aid of the 
Japanese Ministry of Education (Nos.17740171).
The work of AS was supported in part by US DOE Contract No.
DE-AC02-98CH10886.

\end{document}